\documentstyle[12pt,aaspp4]{article}

\newcommand{\msun}{$M/M_{\odot}\,$}

\begin{document}
 
\title{Theoretical Models for Classical Cepheids: IV. Mean Magnitudes 
and Colors and the Evaluation 
of Distance, 
Reddening and Metallicity.}
 
\author{Filippina Caputo\altaffilmark{1}, Marcella Marconi\altaffilmark{1}, 
and Vincenzo Ripepi\altaffilmark{1}} 

\lefthead{Caputo, Marconi \& Ripepi}
\righthead{Theoretical Models for Classical Cepheids: III.}

\altaffiltext{1}{Osservatorio Astronomico di Capodimonte, Via Moiariello 16,
80131 Napoli, Italy; caputo@astrna.na.astro.it, marcella@cerere.na.astro.it, 
ripepi@cerere.na.astro.it}

\begin{abstract}

We discuss the metallicity effect on the theoretical visual and 
near-infrared Period-Luminosity ($PL$)
and Period-Luminosity-Color ($PLC$) relations of classical Cepheids, 
as based on nonlinear, nonlocal and time--dependent
convective pulsating models at varying chemical composition. 
In view of the two usual methods of averaging (magnitude-weighted and
intensity-weighted) observed magnitudes and colors 
over the full pulsation cycle, 
in the first part of the paper we 
briefly discuss the differences between static and mean quantities. 
We show that the behavior of the synthetic 
mean magnitudes and colors 
fully reproduces the 
observed trend of Galactic Cepheids, supporting the validity
of the model predictions.
In the second part of the paper we show how the estimate 
of the mean reddening $E_{B-V}$ and true distance modulus 
$\mu_0$ of a galaxy from  
Cepheid $VK$ photometry 
depend on the adopted metal content,   
in the sense that larger metallicities drive the host galaxy to 
lower extinctions and distances. Conversely, 
self-consistent estimates of the Cepheid mean reddening, distance 
and metallicity may be derived if three-filter 
data (in this paper $BVK$) are taken into account. 
By applying the theoretical $PL$ and $PLC$ relations to available $BVK$ 
data of Cepheids in the Magellanic Clouds we eventually obtain 
$Z\sim$0.008, $E_{B-V}\sim$ 0.02 mag, 
$\mu_0\sim$ 18.63 mag for LMC and 
$Z\sim$0.004, $E_{B-V}\sim$ 0.01 mag, 
$\mu_0\sim$ 19.16 for SMC. The 
discrepancy between such reddenings and the current values 
based on $BVI$ data ($E_{B-V}\sim$ 0.074 mag [LMC]  
and $\sim$ 0.054 mag [SMC]) is briefly 
discussed. 

\end{abstract}
 
\noindent
{\em Subject headings:} Cepheids --- Galaxy: stellar content ---
Magellanic Clouds  --- stars: distances --- stars: oscillations

\pagebreak
\section{INTRODUCTION}
 
\noindent
Cepheids are the best "standard candles" on 
extragalactic distance scales because 
they are bright and easily identified, and, 
most importantly, because the physics of their variability 
is well understood. From the general considerations of 
the Stefan's law connecting luminosity 
$L$, radius $R$ and effective temperature $T_e$, and of the Ritter's 
equation between the pulsation period $P$ and 
the mean density $\rho$, one derives that $P=f(L,M,T_e)$. 
If it is assumed that stellar evolution 
theory predicts a close relation between mass and luminosity, then 
the natural outcome into the observative plane is a $PLC$ relation where 
the absolute magnitude $M_j$ in a given photometric 
bandpass is a linear function of the period and color index [$CI$], as given by 

$$M_j = \alpha + \beta logP + \gamma [CI].\eqno(1)$$ 

On the other hand, the occurrence of stable pulsation deals only with a 
narrow (but finite!) zone in the HR diagram and the existence itself of 
the "Cepheid instability strip" led to neglect the color term in favour of 
a $PL$ relation which is often given in the linear form

$$\overline{M_j} = a + b logP,\eqno(2)$$

\noindent
where $\overline{M_j}$ is the "statistical" magnitude of 
Cepheids at a given period.  

As beautifully introduced in the pioneering work by Sandage (1958),
Sandage \& Gratton (1963), and Sandage \& Tammann (1968), only 
the $PLC$ relation is able to reproduce the tight correlation among 
the parameters of individual Cepheids. For work on the distance scale, this 
means that any individual Cepheid could be used to get accurate estimates 
of the distance to the parent galaxy. On the contrary, the adoption of a 
$PL$ relation requires a large sample of variables in order to decrease 
the error due to deviations from the statistical ridge line 
(see Cox 1980;
Madore \& Freedman 1991; Tanvir 1998 for details and references). There 
is a further effect which is worth to be considered, namely the metallicity 
sensitivity of the $PL$ and $PLC$ relations. This has been matter of a long 
debate (see Kochanek 1997 for a review on the argument), but all 
the available empirical estimates of the metallicity 
correction seem to depend on the Cepheid sample and on the method
adopted for disentangling reddening from metallicity effects. On the 
theoretical side, only in the very recent time an 
extensive grid of Cepheid nonlinear convective models has become available 
(Bono, Marconi \& Stellingwerf 1999 [Paper I]; 
Bono, Caputo, Castellani \& Marconi 1999 [Paper II]; 
Bono, Castellani \& Marconi 1999 [Paper III]), providing 
us with the predicted locations of both blue and red
edges of the instability strip, as well as with theoretical light curves, 
which are fundamental for properly assessing the 
intrinsic interdependence of the Cepheid pulsational properties. 

In Paper II we presented a set of 
theoretical $PL$ and $PLC$ relations as based on pulsating models with  
four values of  the stellar mass (\msun=5.0, 7.0, 9.0, 11.0) and 
three different chemical compositions
($Y$=0.25, $Z$=0.004; $Y$=0.25, $Z$=0.008; $Y$=0.28, $Z$=0.02),
taken as representative of Cepheids in
the Magellanic Clouds and in the Galaxy. In that paper the pulsator 
bolometric light curves were transformed into blue, 
visual and near-infrared magnitudes and the
intensity-weighted averages 
over a full pulsation cycle (hereafter 
$<M_B>$, $<M_V>$ and $<M_K>$) 
were obtained, together with the colors $<B>$--$<V>$ and $<V>$--$<K>$. 
On this basis, theoretical Period-Luminosity and Period-Luminosity-Color 
relations (hereafter $<PL>$ and $<PLC>$ for the intensity-weighted averages) 
were given for the various metallicities, as derived by least squares 
solutions through the models.  
 
Along this line, we present in Section 2 similar relationships which are 
based on the synthetic magnitude-weighted quantities 
$(M_V)$, $(M_K)$, $(B-V)$, and $(V-K)$. This allows a comparison 
of our synthetic mean 
magnitudes and colors with the observed trends of Galactic Cepheids. 
In Section 3 we analyse the metallicity effect on 
the predicted $PL$ and $PLC$ 
relations and we discuss the possibility to disentangle reddening from 
metallicity effects by using three-filter photometric data.  
The last section contains a summary of the main results. 

\section{Mean Magnitudes and Colors}

\noindent
As a first step of our analysis, 
the bolometric light curve provided by nonlinear, nonlocal and 
time-dependent convective models has been 
transformed into the observative plane by means of the 
grid of static 
atmosphere models computed by Castelli, Gratton \& Kurucz (1997a,b). 
Then, the predicted aforementioned mean visual 
$(M_V)$ and $<M_V>$ magnitudes
are derived. As shown in 
Fig. 1 for the case $Z$=0.02, none out of the two means is equal to 
$M_{V,stat}$, the "static" magnitude the star would have were it 
not pulsating, over the whole range of period. Moreover, 
Fig. 2 shows that the difference ($M_V$)--$<M_{V}>$ is 
always positive and may amount up to 0.10 mag, depending on the shape of the
light curve. A straightforward test of these theoretical results
is presented in Fig. 3 where the 
observed trend of Galactic Cepheids (data by Coulson, Caldwell 
\& Gieren 1985; Coulson \& Laney 1985; Moffett \& Barnes 1985) is shown. 
As for the near-infrared magnitudes, we obtain that 
the discrepancy between ($M_K$) and $<M_{K}>$ 
is always quite insignificant ($\le$ 0.01 mag). 

Passing to the colors, we find again some significative differences between 
static and mean values (see Fig. 4 for the case $Z$=0.02), with the predicted 
($B-V$) and ($V-K$) colors always redder than   
$<B>-<V>$ and $<V>-<K>$ (see Fig. 5 for the case $Z$=0.02). Also in 
this case the predicted trend is in fully agreement with the observed data 
of Galactic Cepheids (Fig. 6). In our belief such a concordance between 
predictions and observations supports the validity of our models 
and, in the same time, suggests some 
caution against theoretical relations based on static magnitudes and colors.  

Given the difference between intensity-weighted and magnitude-weigthed 
values, we expect that the predicted $PL$ 
and $PLC$ relations for a fixed chemical composition 
would sligthly depend on the adopted definition of the pulsator mean magnitude 
and color. For this reason, following the procedure presented in Paper II, 
we derive the quadratic visual $(PL_V)$ relation and the linear 
near-infrared $(PL_K)$ relation for magnitude-weighted means. Table 1 
summarizes the $PL$ relations derived here and in Paper II. 
As shown in Fig. 7, where the mean visual magnitude $<M_V>$ of 
fundamental pulsators\footnote{The distribution of pulsators 
in the period-magnitude plane 
is constrained by the calculated blue and red edges of the instability 
strip at varying mass and luminosity (see Paper II).} is plotted 
against period, the  
quadratic $PL_V$ relations (see lower 
panel) are demanded from the pulsator distribution 
in the log$P$-$<M_V>$ plane. Wishing 
to constrain the results into reliable linear approximations, 
we need to "break" the $PL_V$ relations at log$P\sim$ 1.4 
(see upper panel of Fig. 7), with the coefficients as listed 
in Table 2.  As for the intrinsic scatter of
the $PL$ relations, 
the results in Paper II are confirmed, namely
$\sigma_{V,PL}\sim$0.26 mag (quadratic solution), 
$\sigma_{V,PL}\sim$ 0.12 mag (linear solutions) and 
$\sigma_{K,PL}\sim$0.12 mag (linear solution), at fixed 
chemical composition. Note that all 
these results are based on the hypothesis of uniformly populated 
instability strip. Thus, since the $PL$ is a "statistical" relation, both 
the slope and zero-point (and the intrinsic dispersion, as well) will 
obviously depend 
on the pulsator distribution within the instability strip, mainly at the 
shorter wavelengths. 

Concerning the period-luminosity-color relationships, Table 3 gives 
the $\alpha$, $\beta$, and $\gamma$ coefficients of 
the $PLC_V$ relation correlating visual magnitude and 
$B-V$ colors and of the $PLC_K$ relation correlating visual magnitude and
$V-K$ colors, as derived by least square solutions through
our fundamental models at fixed metallicity. As already shown in 
Paper II, the intrinsic scatter of these relations is very small,
namely $\sigma_{V,PLC}\sim$0.05 mag.

As a whole, besides a clear correlation with the chemical 
composition, all the predicted $PL$ and $PLC$ 
relations are slightly dependent on the the adopted way of averaging 
magnitudes and colors. It follows that, if one is comparing observations with 
theoretical relations, then the same type of mean should be used, otherwise 
some systematic error arises on the derived distance modulus. Even though 
this doesn't concern skilled astronomers, we wish to mention that the 
use of an improper $PL_V$ relation is less dangerous than an 
improper $PLC_V$ relation: in the former case the systematic 
effect on the derived distance modulus is always smaller than the 
intrinsic scatter of the relation, whereas in the latter case it 
amounts up to $\pm$ 0.10 mag which is twice $\sigma_{V,PLC}$. 

\section{The Sensitivity of the $PL$ and $PLC$
Relations}

\noindent
Having established that the predicted $PL$ and $PLC$ relations depend 
on the metallicity, it seems important to show their role in our 
attempts to derive the distance to external galaxies. 
 
It is well known that the Cepheid apparent distance modulus $\mu_j$ 
derived from the observed magnitude $m_j$ and the predicted 
relation $PL_j$ 
is connected to the true modulus $\mu_0$ by the 
simple equation 

$$\mu_{j,PL_j} = \mu_0 + R_j*E(B-V)\eqno(3),$$ 

\noindent
where $E(B-V)$ is the total reddening and $R_j$ is the ratio of the total to 
selective absorption in the $j$-filter. As an example,  
the Cardelli, Clayton \& Mathis (1989) model of the extinction 
curve yields $\mu_{V,PL_V}$=$\mu_0$+3.10$E(B-V)$ and 
$\mu_{K,PL_K}$=$\mu_0$+0.37$E(B-V)$. 

From a theoretical point of view, the difference between the 
visual and near-infrared apparent distance
modulus based on $PL_V$ and $PL_K$ 
should suffice to estimate the mean reddening of 
the host galaxy, and then the true distance. 
The results would slightly depend on the 
adopted values of $R_V$ and $R_K$ (see also Laney \& Stobie 1993; 
Fouqu\`e \& Gieren 1997; Kochaneck 1997), but here we wish to 
emphasize that 
both the zero-point and the slope of the
predicted $PL_V$ relations are 
dependent on the pulsator metallicity and distribution, 
and that the large width of the instability
strip might cause in the $V$-band an intrinsic scatter of 
$\sigma_{V,PL_V}\sim$0.26 mag. Conversely, the spurious 
effect introduced by the
intrinsic width of the instability strip is dramatically decreased in the 
near-infrared $PL_K$ ($\sigma_{K,PL}\sim$0.12 mag) and 
$PLC_K$ relations ($\sigma_{V,PLC}\sim$0.05 mag). This yields that, 
if dealing with only $VK$ measurements, the use of 
the $PL_V$ relation should be avoided and 
better estimates of the mean reddening (and distance, as well) 
should be derived from $PL_K$ and 
$PLC_K$ relations. 

Concerning the distance modulus derived from theoretical $PLC$ relations, 
the resulting reddening sensitivity 
depends on the color term  and, in a minor way, on the adopted 
method of averaging magnitudes and 
colors. As a whole, from Eq. (1) one has 

$$\mu_{V,PLC_j} = \mu_0 + r_j*E_{B-V},\eqno(4)$$

\noindent
where, following Cardelli, Clayton \& Mathis (1989),
$r_V$=(3.10-$\gamma_V$) with $B-V$ colors and  
$r_K$=(3.10-2.73$\gamma_K$) with $V-K$ colors.  

As a conclusion, assuming 
$\Delta_1$ as the difference between $\mu_{V,PLC_K}$ and
$\mu_{K,PL_K}$, one has that Cepheid $VK$ measurements, together with  
predicted $PL_K$ and $PLC_K$ relations, would allow to
estimate the mean reddening from $\Delta_1$=2.73(1-$\gamma_K)E_{B-V}$, 
and then $\mu_0$ of the host galaxy. However, both 
the zero-point and the slope of
$PL_K$ and $PLC_K$ vary with the metallicity, 
even though with a 
less degree in comparison with the visual relationships. 
Thus, this two-filter 
method requires the previous knowledge of the Cepheid chemical 
composition as forcing the slopes to be "universal" 
(f.e., that found for the LMC Cepheids) may yield unreal differences in 
the extinction and true distance modulus. 
The obstacle of unknown metallicities could be overcome if 
three-filter photometry is available, e.g. $BVK$ data. As a fact,   
the information
coming from $PLC_V$ provides the difference
$\Delta_2$ between $\mu_{V,PLC_K}$ and
$\mu_{V,PLC_V}$ and then a new estimate 
of the mean reddening and 
distance from $\Delta_2$=($\gamma_V$-2.73$\gamma_K)E_{B-V}$. 

On the other hand, all the predicted $PL$ and $PLC$ relations depend in 
different ways on the metallicity and the coefficients in Table 1 and 
Table 3 (see also Fig.s 15, 16 and 17 in Paper II) 
suggest that $\mu_{K,PL_K}$ and $\mu_{V,PLC_K}$ should slightly increase with 
decreasing the adopted metal content, whereas smaller $\mu_{V,PLC_V}$ are 
expected at the lower adopted metallicities. Consequently, 
also the dependence of $\Delta_1$ and $\Delta_2$ on the adopted 
metallicity is expected with  
different slope. However, at the Cepheid actual  
metallicity the results from $\Delta_1$ and $\Delta_2$ should  
be coincident and the net result is that $BVK$ photometry 
may provide self-consistent estimates of the Cepheid mean reddening 
and metallicity which are necessary to tie down the galaxy distance. 

In order to test this two-color photometry method, we analyze LMC and SMC 
Cepheids (data\footnote{For our
analysis the unreddened values
listed by Laney \& Stobie (1994) have been "reddened" according to the
$E_{B-V}$ values listed by the same authors and following the procedure
described in Laney \& Stobie (1993).} by Laney \& Stobie 1994), assuming to
be unaware of the galaxy metal content. From 
the predicted $<PL_K>$ and $<PLC_K>$ relation we derive, for 
each adopted metallicity, the 
$E_{B-V,1}$ and $\mu_{0,1}$ values listed in columns 2 and 3 of Table 4. 
One notices that 
the resulting estimates decrease 
at the larger metal abundances, confirming 
that within a two-filter photometry framework it is hard
to disentangle reddening from metallicity effects and that 
additional colors are needed to derive independent 
estimates of reddening and distance (see also Laney \& Stobie
1994; Kochanek 1997). 

As a fact, if $B-V$ colors are available and $<PLC_V>$ is used together 
with $<PLC_K>$, then the resulting $E_{B-V,2}$
and $\mu_{0,2}$ values (columns 4 and 5 of Table 4) 
turn out to increase with increasing the 
adopted metallicity. As shown in Fig. 8 and Fig. 9, the 
equality $E_{1,B-V}$=$E_{2,B-V}$ and
$\mu_{0,1}$=$\mu_{0,2}$ occurs at $Z\sim$0.008 for LMC and 
$Z\sim$0.004 for SMC, in close
agreement with the spectroscopic analysis of Luck et al. (1998). 
At the same time we get for LMC 
$E_{B-V}\sim 0.02\pm$ 0.08 mag and $\mu_0\sim 18.63\pm$0.12 mag, and 
for SMC 
$E_{B-V}\sim 0.01\pm$ 0.08 mag and $\mu_0\sim 19.16\pm$0.17 mag.  

The derived true distance moduli appear 
in close agreement with recent estimates (see Walker 1998 and 
references therein) but 
the mean reddening based on theoretical $PL_K$, $PLC_V$ and 
$PLC_K$ relations appears somehow smaller than the current values  
($E_{B-V}\sim 0.074\pm$0.037 mag and 
$E_{B-V}\sim 0.054\pm$0.026 mag for LMC and SMC, respectively) 
found by Caldwell \& Coulson (1985) from 
Cepheid $BVI$ photometry. 

On this subject, let us briefly summarize some important points: 
\begin{enumerate}
\item	Starting from magnitude means of 
simultaneous $\overline{(B-V)}$ and $\overline{(V-I)}$ pairs for 
Galactic Cepheids with known reddening, Dean, 
Warren \& Cousins (1978) obtain the intrinsic locus in the $(B-V)$-$(V-I)$ 
plane as a quadratic relation 
$$(B-V)_0=a+b(V-I)_0+c(V-I)_0^2,$$
\noindent 
giving three solutions, namely $(a,b,c)$=(0.00, 0.58, 0.57), 
$(a,b,c)$=(0.09, 0.30, 0.77) and $(a,b,c)$=(-0.09, 0.72, 0.40). As 
shown in Fig. 10, our synthetic mean colors $(B-V)$ and $(V-I)$ with $Z$=0.02 
fit the Dean, Warren \& Cousins (1978) results, even 
though they show to move from the first two solutions towards the 
third one at $(V-I)_0\ge$ 1.0 mag.    
\item	From the $\overline{(B-V)}$, $\overline{(V-I)}$ 
simultaneous means of LMC and SMC  
Cepheids, and 
adopting the first solution for the intrinsic line, 
Caldwell \& Coulson (1985) calculate 
the Cepheid individual reddening under various assumptions of the 
metal deficiency $D=Z_{Gal}/Z_{MC}$. Adopting $D_{LMC}$=1.4 and 
$D_{SMC}$=4, they derived $E_{B-V}\sim$0.074 mag 
and $E_{B-V}\sim$0.054 mag, respectively. 
However, if $Z_{Gal}\sim$0.02
(Fry \& Carney 1997) and according to Fry \& Carney (1997) results 
($<Z>_{LMC}\sim$0.008, $<Z>_{SMC}\sim$0.004), one has $D_{LMC}$=2.5 and 
$D_{SMC}$=5. Consequently the $BVI$ reddening should become 
$E_{B-V}\sim$0.042 mag [LMC] and $E_{B-V}\sim$0.047 mag [SMC].
\item	Moreover, if the  $(a,b,c)$=(-0.09, 0.72, 0.40) solution by Dean,
Warren \& Cousins (1978) were adopted, then the $BVI$ reddenings 
should further decrease. This is 
clearly shown in Fig. 10 where the Magellanic 
Clouds Cepheids are plotted in comparison with 
the $(a,b,c)$=(0.00, 0.58, 0.57; solid line) and 
$(a,b,c)$=(-0.09, 0.72, 0.40; dashed line) metal-deficient 
unreddened locus. With reference to the solid line the derived 
mean reddenings are $E_{B-V}\sim$0.042 mag [LMC] 
and $E_{B-V}\sim$0.047 mag [SMC], whereas if the dashed line is 
adopted, then also negative extinctions are derived. 
\end{enumerate}

\noindent 
In conclusion, our reddenings based on the theoretical $PL_K$, $PLC_V$ and
$PLC_K$ relations are in qualitative 
agreements with those based on the Cepheid $BVI$ color-color diagram, all 
the results suggesting that the mean reddening in the Magellanic Clouds  
is "low" ($\le$ 0.05 mag). On this basis, the above predicted relations 
for $BVK$ photometry appear adequate to estimate the Cepheid mean metallicity 
and reddening and to get the distance to the host galaxy. 

\section{Summary}

\noindent
Nonlinear, nonlocal and time-dependent 
convective pulsating models of Cepheids 
are used to derive the synthetic intensity-weighted and 
magnitude-weighted mean magnitudes and colors and to predict the visual 
$V$ and near-infrared $K$ period-luminosity and period-luminosity-color 
relations. 
We show that, besides the metallicity effect already presented in Paper II, 
the predicted $PL$ and $PLC$ are slightly dependent on the 
kind of mean values we are dealing with. 
Due to the different values of the color term, 
also the reddening sensitivity of the predicted $PLC$ relations varies with 
both the metallicity of pulsators 
and, in a minor way, the adopted mean magnitudes and colors. The predicted 
differences between the two synthetic mean magnitudes and colors   
are in full agreement with the observed trends of Galactic Cepheids, 
supporting the validity of our model predictions.
 
As regards to the use of $PL$ and $PLC$ relations to get reddening and 
distance of the host galaxy, we show how a two-filter photometry 
(here $VK$ but likely also $VI$, as in the $HST$ Key 
Project) yields $E_{B-V}$ and    
$\mu_0$ which decrease with increasing the adopted metallicity.  
Conversely, if Cepheid data in three bands (e.g. $BVK$) are available, 
then we have the possibility to disentangle 
reddening from metallicity effects. This 
confirms early suggestions (e.g., Laney \& Stobie 1994)   
that a minimum of two-color photometry is needed 
to get independent and reliable estimates of reddening, true distance modulus  
and metallicity. 

From the comparison of the predicted relations 
with published $BVK$ data of 
Cepheids in the Magellanic Clouds we derive $Z\sim$0.008, 
$E_{B-V}\sim$0.02 mag and $\mu\sim$18.63 mag in LMC, and 
$Z\sim$0.004,
$E_{B-V}\sim$0.01 mag and $\mu_0\sim$19.16 mag in SMC. 
The small discrepancy with 
the current reddening values based on 
$BVI$ photometry is briefly discussed, showing that all the 
methods are in qualitative agreements within the errors. 

\noindent
{\it Acknowledgements.} 
We deeply thank the anonymous referees for the valuable comments 
which improved the first version of the paper. 
\pagebreak

\pagebreak

\figcaption [] {The difference between the static magnitude and 
the two different mean magnitudes for fundamental pulsators 
with $Z$=0.02 and different masses.}

\figcaption [] {The difference between the two mean magnitudes at the 
various metallicities. Symbols as in Fig. 1.}

\figcaption [] {As in Fig. 2, but for observed Cepheids in the Galaxy.}

\figcaption [] {The difference between the static $B-V$ color and
the two different means for fundamental pulsators
with $Z$=0.02. Symbols as in Fig. 1.}

\figcaption [] {The difference between the two mean visual colors and 
near-infrared colors, with 
$Z$=0.02. Symbols as in Fig. 1.}

\figcaption [] {As in Fig. 5, but for the observed visual colors of 
Cepheids in the Galaxy.}

\figcaption [] {The log$P$-$<M_V>$ distribution of 
fundamental pulsators with different masses and metallicities, in 
comparison with the predicted quadratic (lower panel) and 
linear (upper panel) $PL_V$ relations.} 

\figcaption [] {Reddening and true distance modulus of LMC as a 
function of metallicity.}

\figcaption [] {Reddening and true distance modulus of SMC as a
function of metallicity.}

\figcaption [] {Fundamental pulsators with $Z$=0.02 (dots) in comparison with 
the intrinsic locus $(B-V)$-$(V-I)$ of Galactic Cepheids. The three lines 
refer to the labelled solutions of the quadratic relation (see text).}

\figcaption [] {(lower panel) - LMC Cepheids in comparison 
with the intrinsic lines of Fig. 10 shifted according to a metal 
deficiency $D$=2.5; (upper panel) - SMC Cepheids in comparison
with the intrinsic lines of Fig. 10 shifted according to a metal
deficiency $D$=5.0. All the data are from Caldwell \& Coulson (1985).} 


\begin{deluxetable}{cccccccc}
\tablecaption{$PL$ relations for fundamental pulsators}
\tablehead{
\colhead{Z\tablenotemark{a}} &
\colhead{a\tablenotemark{b}} &
\colhead{b\tablenotemark{c}} &
\colhead{c\tablenotemark{d}} &
\colhead{a\tablenotemark{e}} &
\colhead{b\tablenotemark{f}} &
\colhead{c\tablenotemark{g}}}
\startdata  
\multicolumn{7}{c}{$ \overline{M_V}=a+b{\log{P}}+c{(\log{P})^2}$} \nl
      &         &     &  &         &         &         \nl
0.004 &-0.67 & -4.39 & +0.76 &-0.75 &-4.21 &+0.66\nl
      &$\pm$0.02 &$\pm$0.13&$\pm$0.06 &$\pm$0.06 &$\pm$0.05&$\pm$0.02\nl
0.008 &-0.84 & -3.92 & +0.63 &-0.84 &-3.97 &+0.63\nl
      &$\pm$0.01 &$\pm$0.06&$\pm$0.03 &$\pm$0.03&$\pm$0.02&$\pm$0.02 \nl
0.02 &-1.48 & -2.53 & +0.21 &-1.37 &-2.84&+0.33 \nl
      &$\pm$0.06 &$\pm$0.23&$\pm$0.13 &$\pm$0.06&$\pm$0.06& $\pm$0.02 \nl
      &          &     & &    &         &         \nl
  \multicolumn{7}{c}{$ \overline{M_K}=a+b\log{P} $} \nl
      &         &        & &        &         &         \nl
0.004 &-2.67 & -3.26 & -- &-2.66 &-3.27 & -- \nl
      &$\pm$0.09 &$\pm$0.10& -- &$\pm$0.08 &$\pm$0.09 & -- \nl
0.008 &-2.69 & -3.17 & -- &-2.68 &-3.19 & --\nl
      &$\pm$0.09 &$\pm$0.10& -- &$\pm$0.08 &$\pm$0.09 & --\nl
0.02  &-2.73 & -3.03 & -- &-2.73 &-3.03 & --\nl
      &$\pm$0.06 &$\pm$0.06& -- &$\pm$0.07 &$\pm$0.07 & -- \nl
\enddata
\tablenotetext{a}{Metal content.
\hspace*{0.5mm} $^b$ Zero point of the ($PL_j$) relation.
\hspace*{0.5mm} $^c$ Logarithmic period coefficient of the ($PL_j$) relation.
\hspace*{0.5mm} $^d$ Squared logarithmic period coefficient 
of the ($PL_j$) relation.
\hspace*{0.5mm} $^e$ Zero point of the $<PL_j>$ relation. 
\hspace*{0.5mm} $^f$ Logarithmic period coefficient of the $<PL_j>$ relation. 
\hspace*{0.5mm} $^g$ Squared logarithmic period coefficient 
of the $<PL_j>$ relation.}
\end{deluxetable}

\begin{deluxetable}{ccccc}
\tablecaption{Linear $PL_V$ relations for fundamental 
pulsators}
\tablehead{
\colhead{Z\tablenotemark{a}} &
\colhead{a\tablenotemark{b}} &
\colhead{b\tablenotemark{c}} &
\colhead{c\tablenotemark{d}} &
\colhead{a\tablenotemark{e}} 
}
\startdata
  \multicolumn{5}{c}{$ \overline{M_V}=a+b\log{P} $} \nl
               &        &        &         &         \nl
 \multicolumn{5}{c}{log$P <$ 1.4  } \nl
	  &        &        &         &         \nl
0.004 & -1.21 & -3.02 & -1.20 & -3.04 \nl
      &$\pm$0.03&$\pm$0.05&$\pm$0.03&$\pm$0.05\nl
0.008 & -1.33 & -2.75 & -1.32 & -2.79 \nl
      &$\pm$0.05&$\pm$0.06&$\pm$0.04&$\pm$0.06\nl
0.02  & -1.64 & -2.14 & -1.62 & -2.22 \nl
      &$\pm$0.03&$\pm$0.04&$\pm$0.03&$\pm$0.04\nl
      &         &        &        &            \nl
 \multicolumn{5}{c}{ log$P >$ 1.4  } \nl
	 &        &        &         &         \nl
0.004 & -2.39 & -2.08 & -2.42 & -2.10 \nl
      &$\pm$0.01&$\pm$0.04&$\pm$0.02&$\pm$0.04\nl
0.008 & -2.38 & -1.94 & -2.47 & -1.92 \nl
      &$\pm$0.05&$\pm$0.07&$\pm$0.04&$\pm$0.06\nl
0.02  & -2.30 & -1.71 & -2.59 & -1.56 \nl
      &$\pm$0.03&$\pm$0.04&$\pm$0.04&$\pm$0.05\nl
\enddata
\tablenotetext{a}{Metal content.
\hspace*{0.5mm} $^b$ Zero point of the $(PL_V)$ relation.
\hspace*{0.5mm} $^c$ Logarithmic period coefficient of the $(PL_V)$ relation.
\hspace*{0.5mm} $^d$ Zero point of the $<PL_V>$ relation.
\hspace*{0.5mm} $^e$ Logarithmic period coefficient of the $<PL_V>$ relation.}
\end{deluxetable}
\begin{deluxetable}{cccccccc}
\tablecaption{$PLC$ relations for fundamental pulsators}
\tablehead{
\colhead{Z\tablenotemark{a}} &
\colhead{$\alpha$\tablenotemark{b}} &
\colhead{$\beta$\tablenotemark{c}} &
\colhead{$\gamma$\tablenotemark{d}}}
\startdata
  \multicolumn{4}{c}{$(M_V)$=$\alpha$+$\beta$log$P$+$\gamma$($B-V$)} \nl
0.004 &-2.58 & -3.72 & +3.05\nl
      &$\pm$0.04&$\pm$0.04&$\pm$0.07\nl
0.008 &-2.78 & -3.76 & +3.20\nl
      &$\pm$0.05&$\pm$0.05&$\pm$0.09\nl
0.02  &-3.27 & -3.86 & +3.60\nl
      &$\pm$0.08&$\pm$0.13&$\pm$0.24\nl
      &                   &         &         \nl
  \multicolumn{4}{c}{$<M_V>$=$\alpha$+$\beta$log$P$+$\gamma$[$<B>-<V>$]} \nl
0.004 &-2.54 & -3.52 & +2.79\nl
      &$\pm$0.04&$\pm$0.03&$\pm$0.07\nl
0.008 &-2.63 & -3.55 & +2.83\nl
      &$\pm$0.04&$\pm$0.03&$\pm$0.06\nl
0.02  &-2.98 & -3.72 & +3.27\nl
      &$\pm$0.07&$\pm$0.10&$\pm$0.18\nl
      &                   &         &         \nl
    \multicolumn{4}{c}{$(M_V)$=$\alpha$+$\beta$log$P$+$\gamma$($V-K$)} \nl
0.004 &-3.46 & -3.63 & +1.65\nl
      &$\pm$0.04&$\pm$0.04&$\pm$0.04\nl
0.008 &-3.40 & -3.62 & +1.62\nl
      &$\pm$0.04&$\pm$0.04&$\pm$0.04\nl
0.02  &-3.27 & -3.54 & +1.53\nl
      &$\pm$0.05&$\pm$0.06&$\pm$0.05\nl
      &                  &         &         \nl
    \multicolumn{4}{c}{$<M_V>$=$\alpha$+$\beta$log$P$+$\gamma$[$<V>-<K>$]} \nl
0.004 &-3.44 & -3.61 & +1.64\nl
      &$\pm$0.04&$\pm$0.03&$\pm$0.03\nl
0.008 &-3.37 & -3.60 & +1.61\nl
      &$\pm$0.04&$\pm$0.03&$\pm$0.03\nl
0.02  &-3.25 & -3.55 & +1.53\nl
      &$\pm$0.04&$\pm$0.05&$\pm$0.04\nl
\enddata
\tablenotetext{a}{Metal content.
\hspace*{0.5mm} $^b$ Zero point.
\hspace*{0.5mm} $^c$ Logarithmic period coefficient.
\hspace*{0.5mm} $^d$ Color coefficient.
}
\end{deluxetable}
\begin{deluxetable}{ccccc}
\tablecaption{Metallicity sensitivity of the 
reddening and true distance 
modulus of the Magellanic Clouds, 
as derived from $PL$ and $PLC$ relations. $E_{B-V,1}$ and 
$\mu_{0,1}$ are from $<PL_K>$ and $<PLC_K>$, while $E_{B-V,2}$ and
$\mu_{0,2}$ come from $<PLC_K>$ and $<PLC_V>$.}
\tablehead{
\colhead{Z\tablenotemark} &
\colhead{$E_{B-V,1}$} &
\colhead{$\mu_{0,1}$} &
\colhead{$E_{B-V,2}$} &
\colhead{$\mu_{0,2}$}}
\startdata
\multicolumn{5}{c}{LMC} \nl   
      &          &         &         &         \nl
0.004 & +0.03 & 18.70 & -0.05 & 18.56 \nl
      &$\pm$0.07&$\pm$0.11&$\pm$0.09&$\pm$0.13\nl
0.008 & +0.01 & 18.63 & +0.02 & 18.64 \nl
      &$\pm$0.07&$\pm$0.11&$\pm$0.09&$\pm$0.13\nl
0.02  & -0.09 & 18.51 & +0.25 & 18.88 \nl
      &$\pm$0.09&$\pm$0.13&$\pm$0.15&$\pm$0.17\nl
      &          &         &         &         \nl
\multicolumn{5}{c}{SMC} \nl
      &          &         &         &         \nl
0.004 & +0.01 & 19.14 & +0.02 & 19.18 \nl
      &$\pm$0.07&$\pm$0.17&$\pm$0.09&$\pm$0.17\nl
0.008 & -0.04 & 19.07 & +0.09 & 19.26 \nl
      &$\pm$0.07&$\pm$0.18&$\pm$0.09&$\pm$0.17\nl
0.02  & -0.14 & 18.94 & +0.47 & 19.62 \nl
      &$\pm$0.09&$\pm$0.19&$\pm$0.15&$\pm$0.19\nl
\enddata
\end{deluxetable}

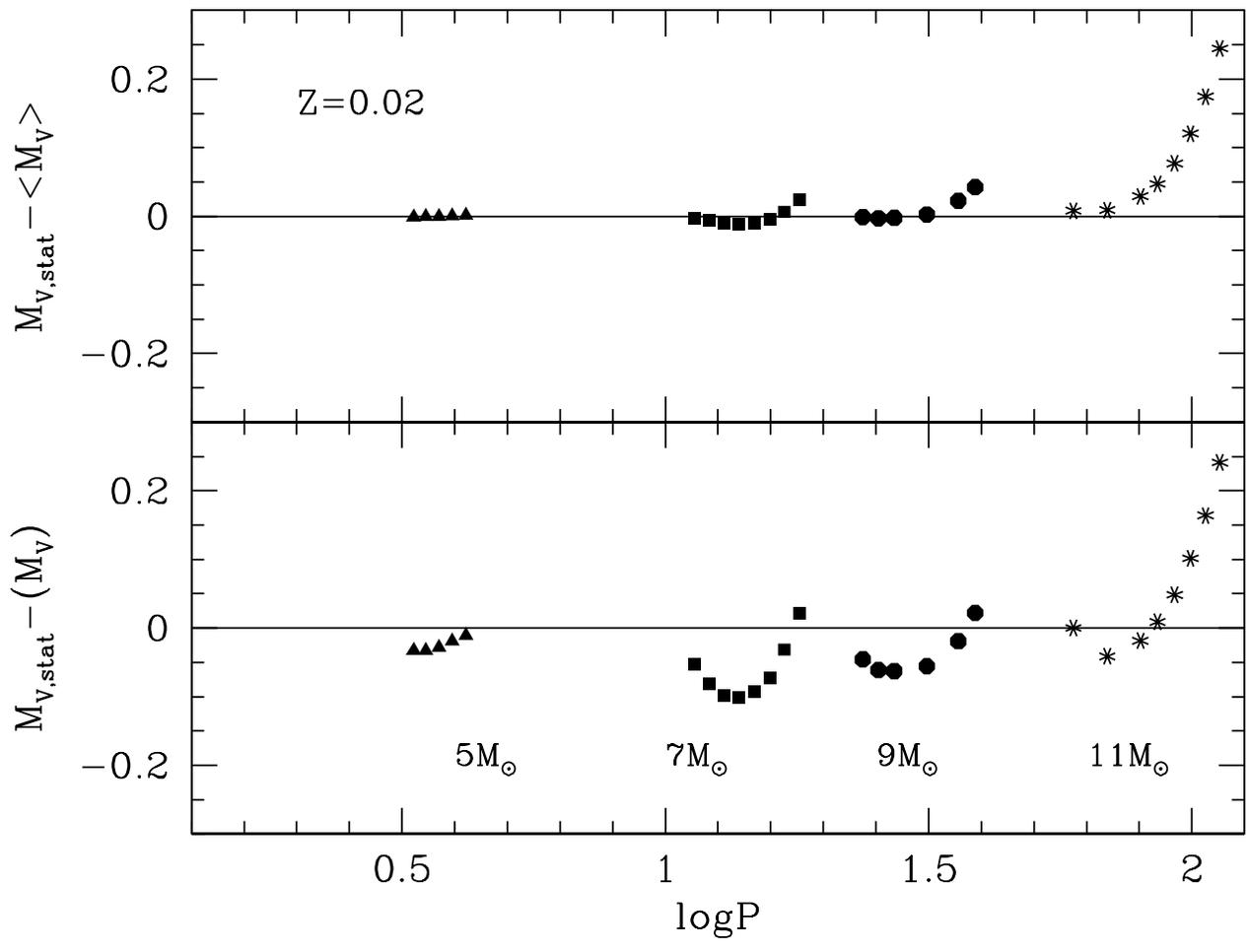
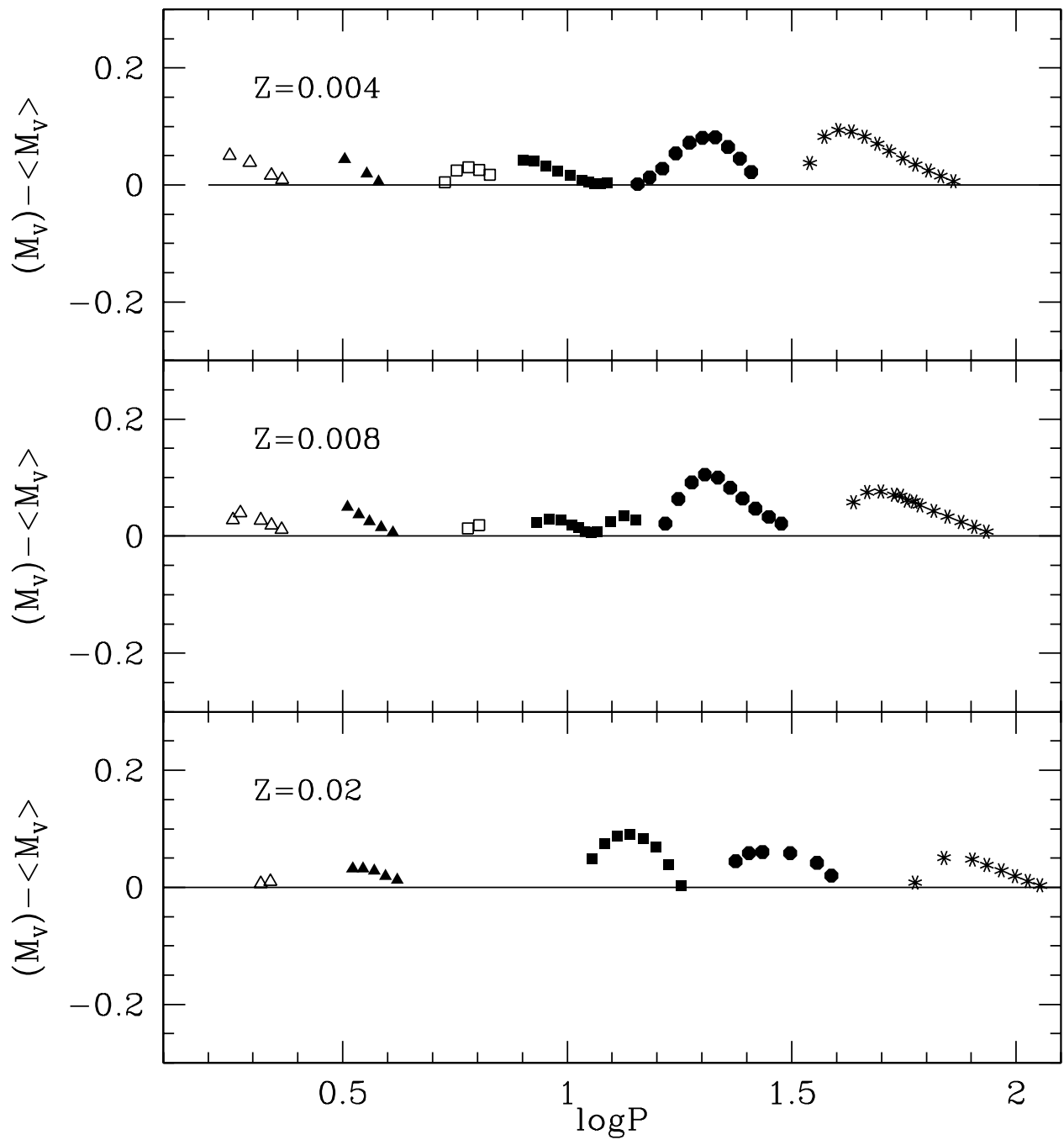
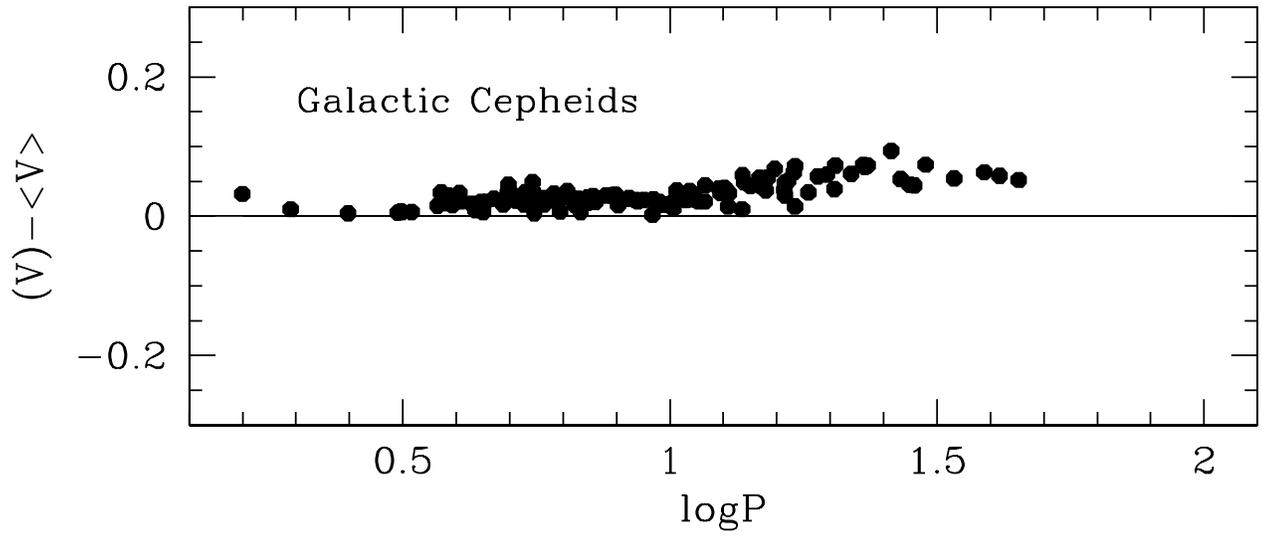
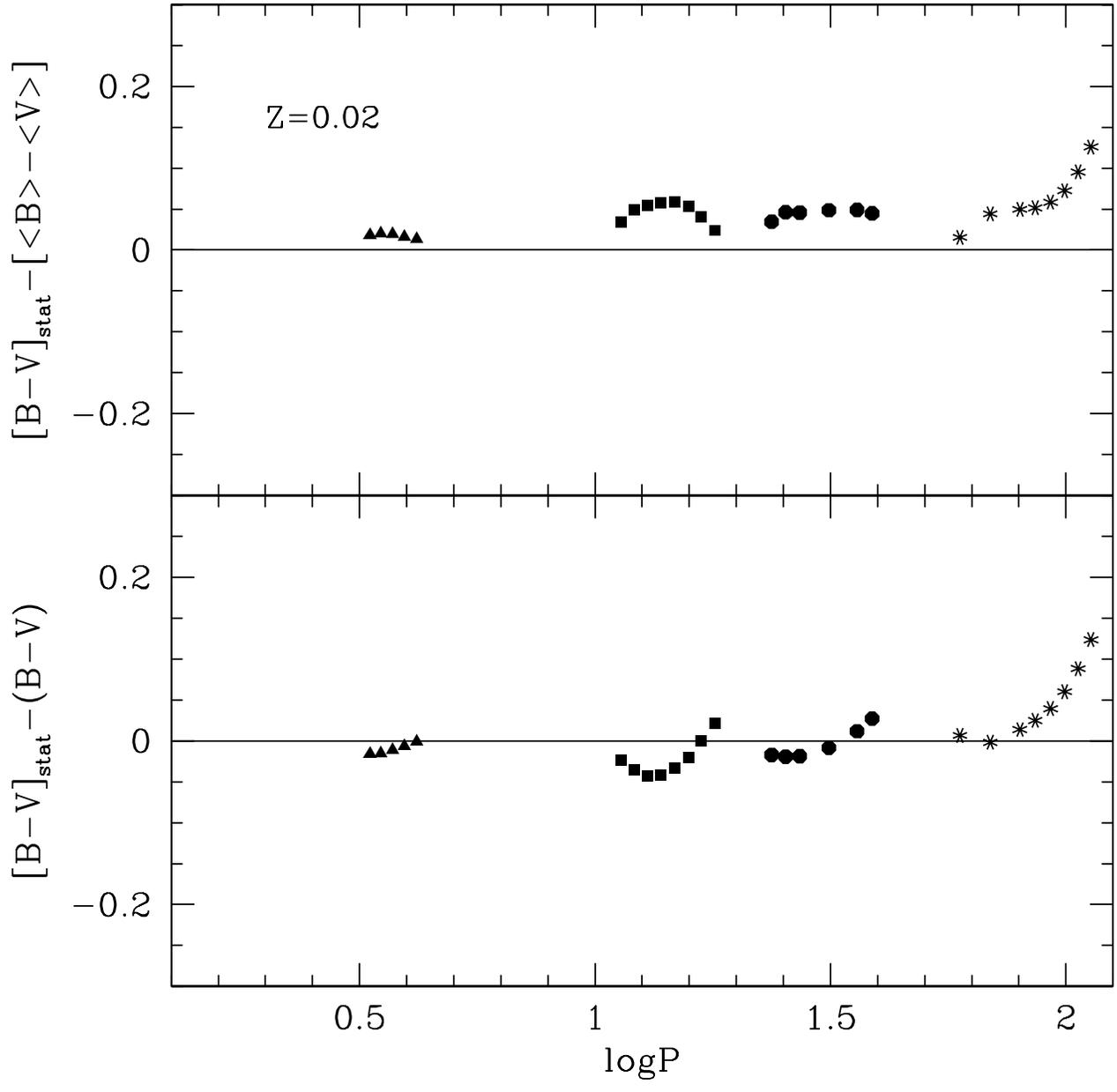
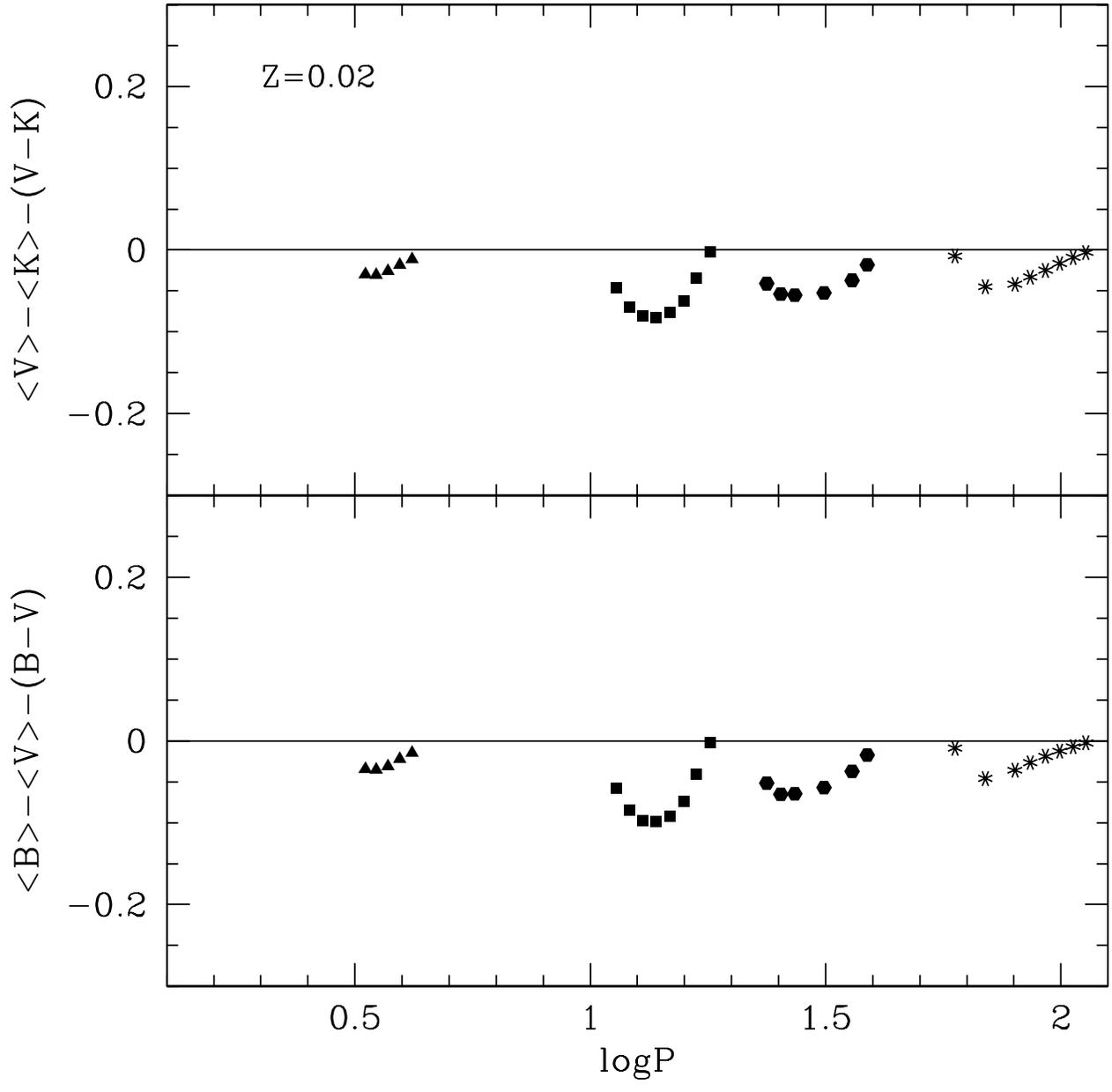
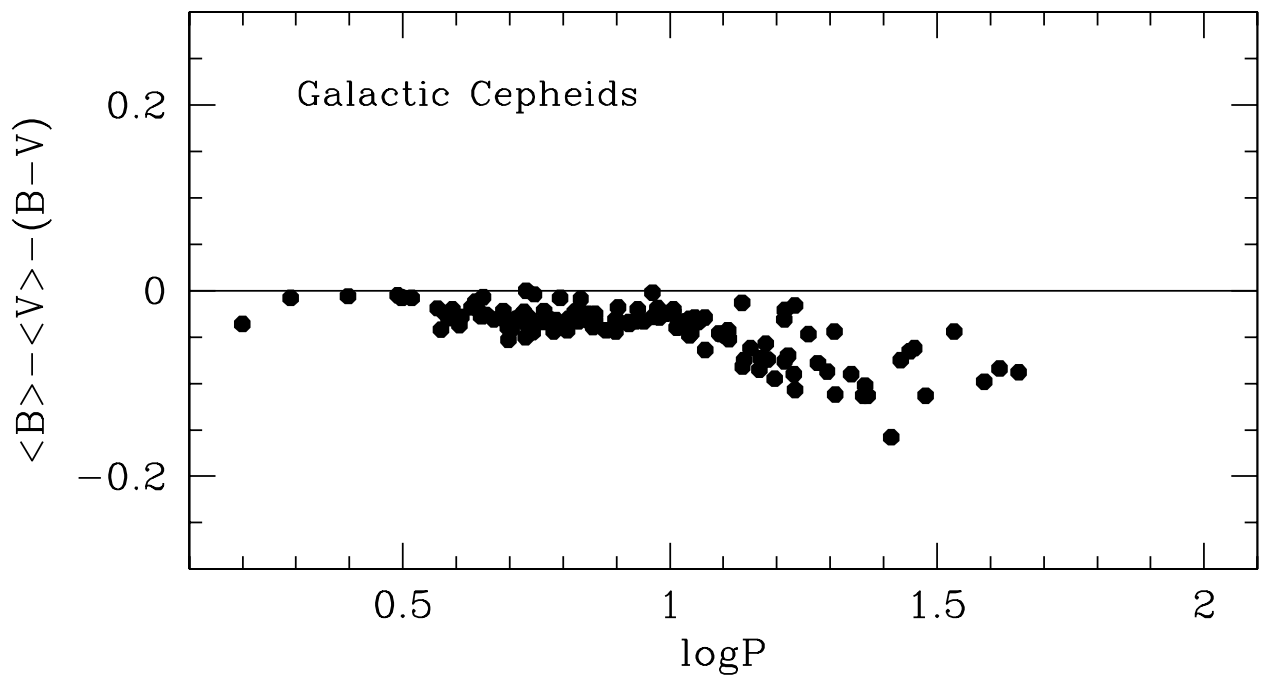
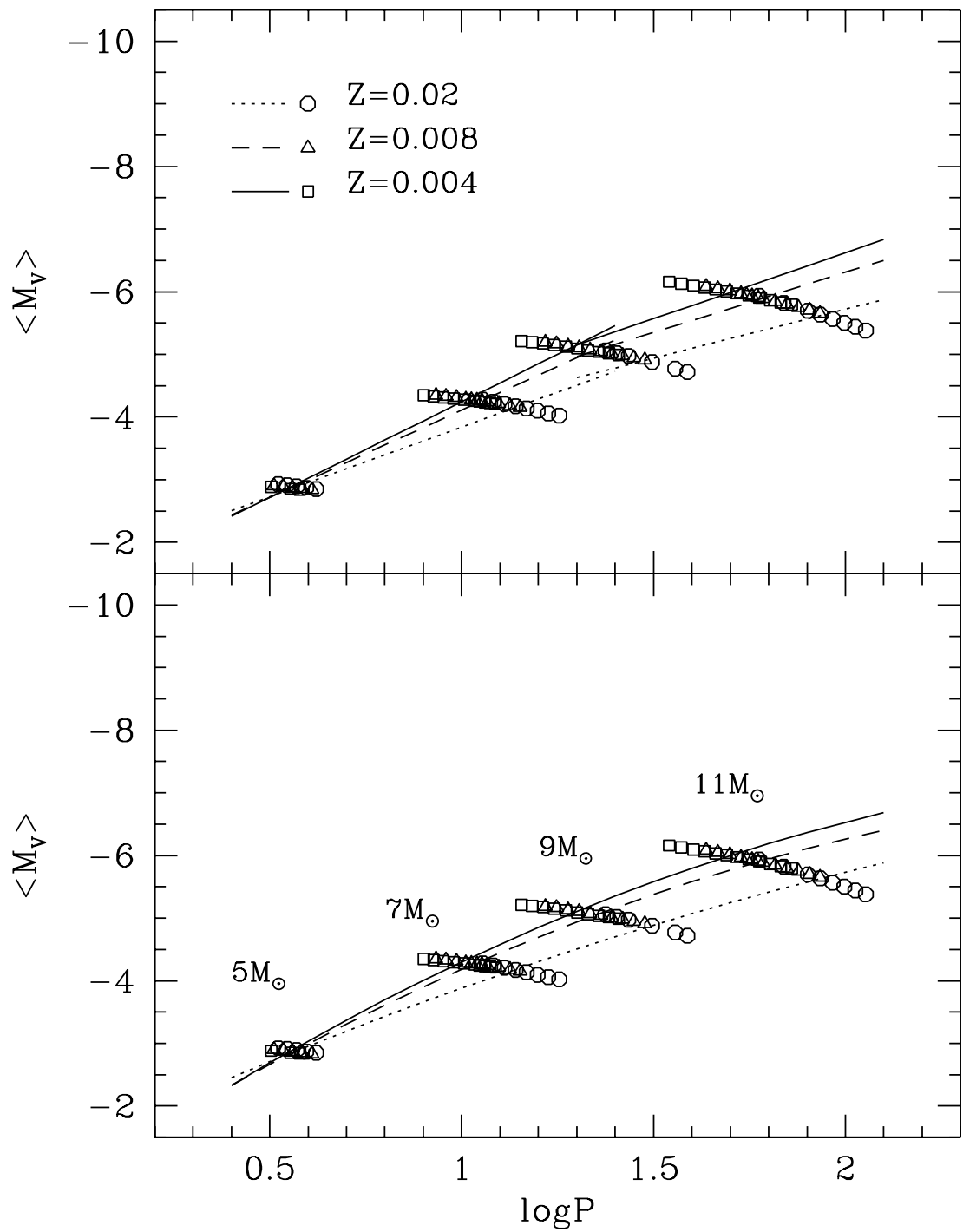
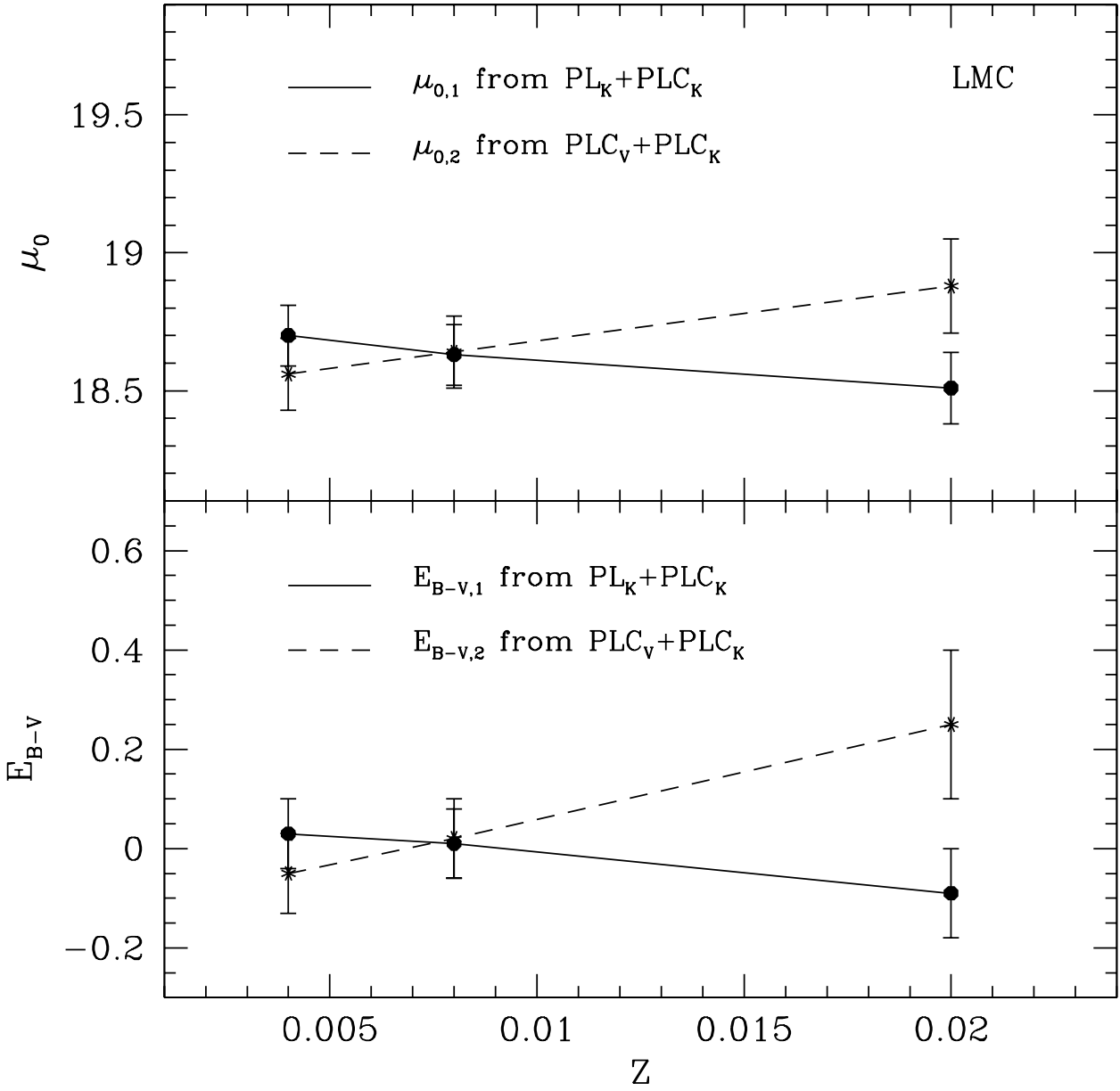
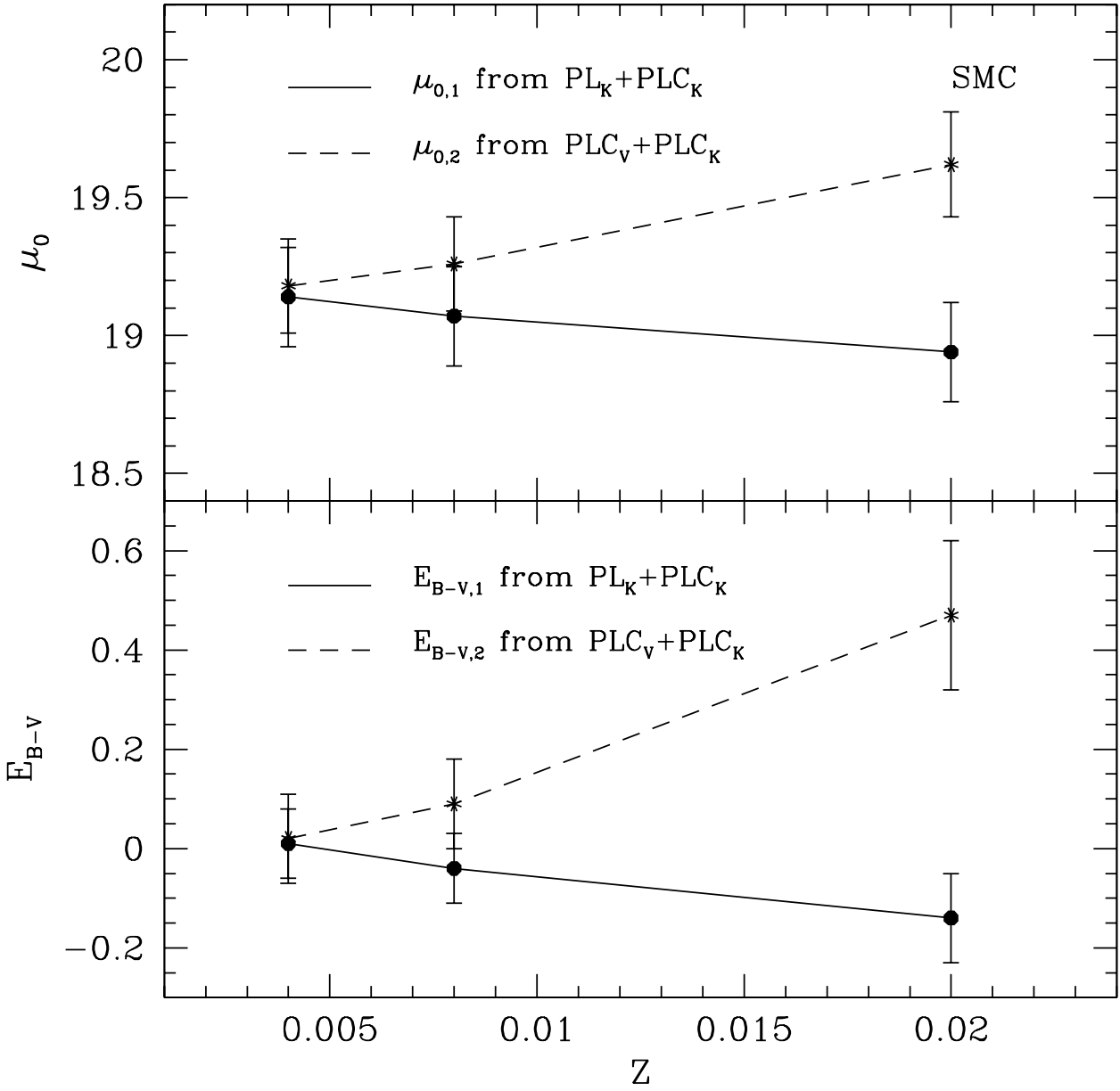
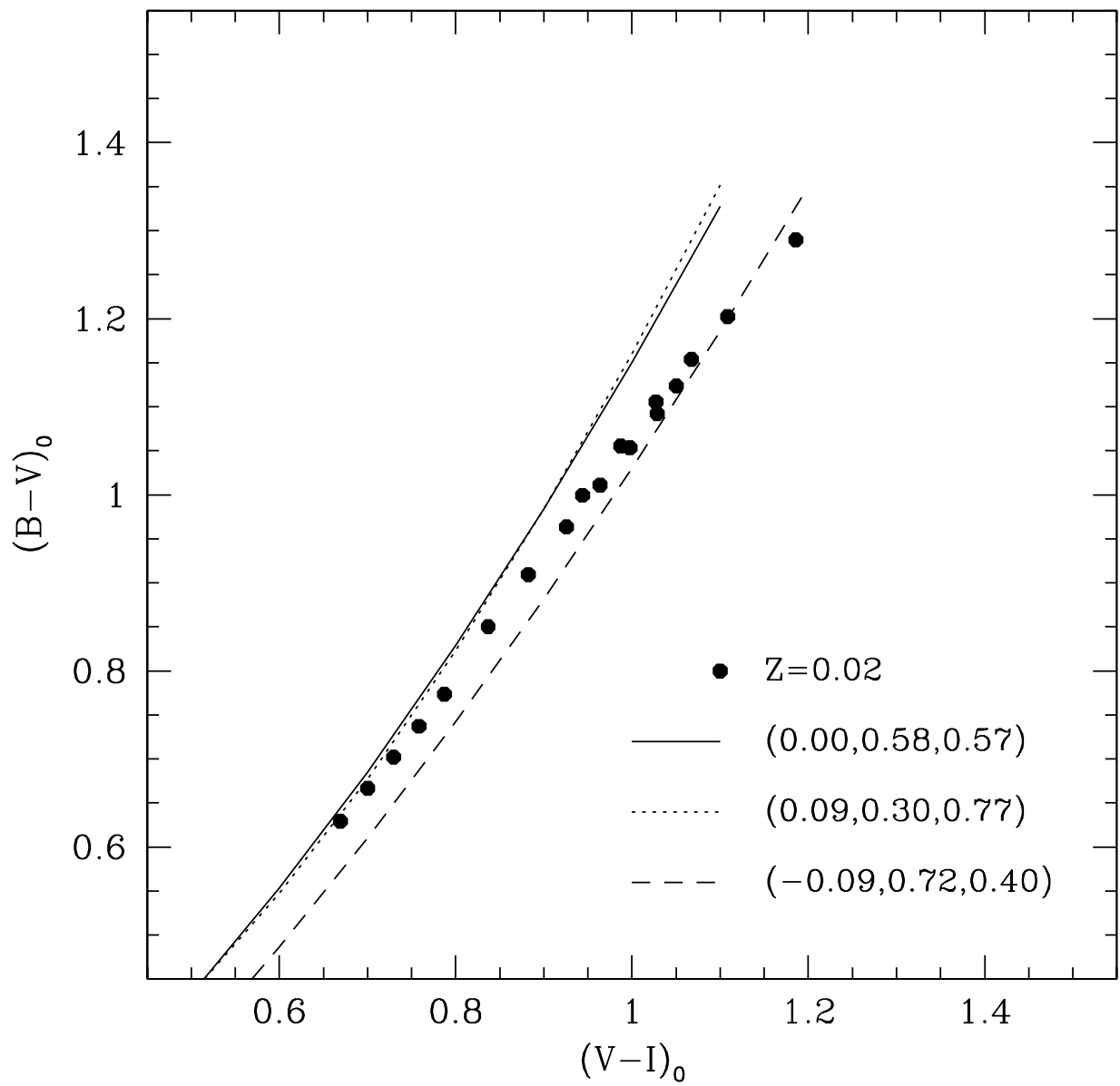
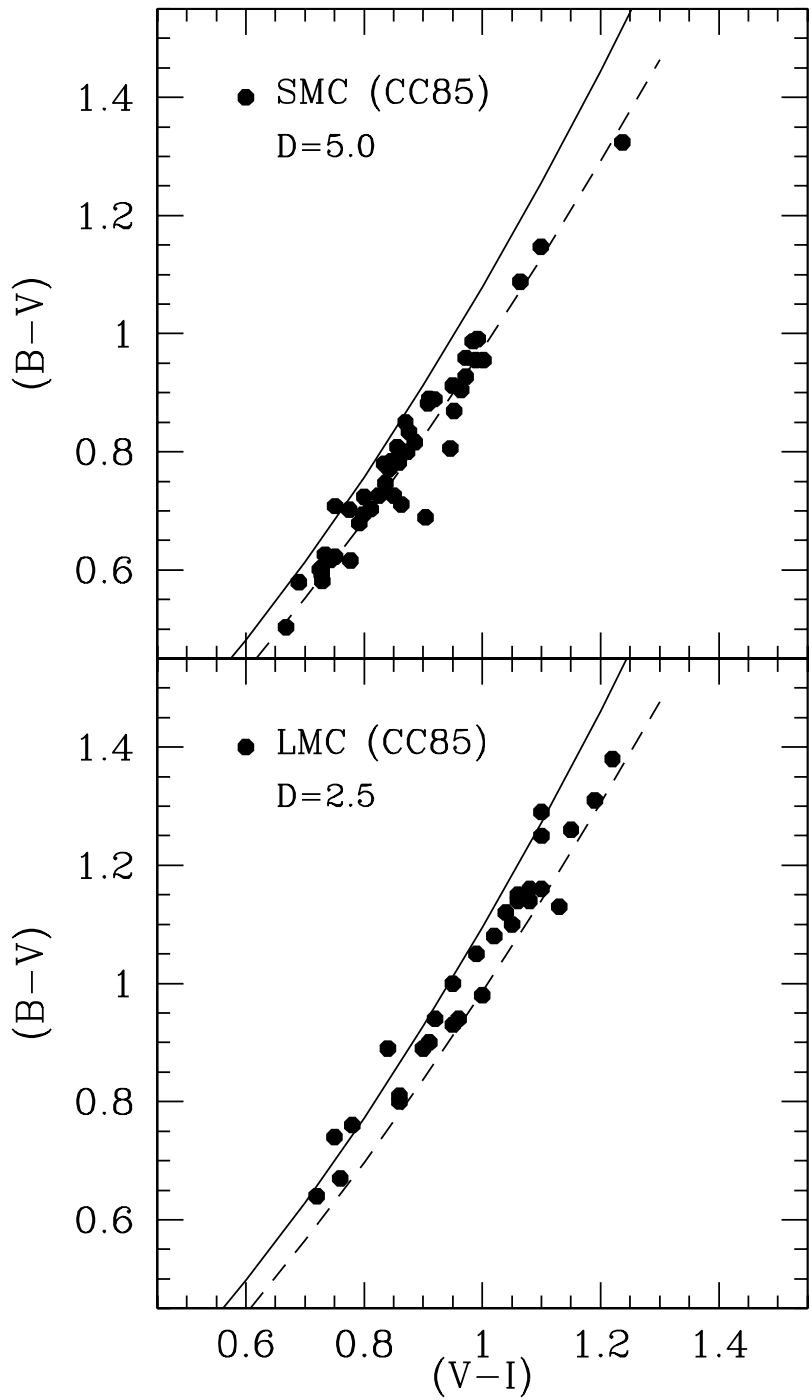
\end{document}